\documentclass[pre,twocolumn]{revtex4-1}

\usepackage{graphicx}
\usepackage{dcolumn}
\usepackage{bm}

\usepackage[utf8]{inputenc}
\usepackage[T1]{fontenc}
\usepackage{mathptmx}

\usepackage{amsmath}
\usepackage{mathtools}
\usepackage{amssymb}

\usepackage{pst-node}
\usepackage{verbatim}   
\usepackage{color}      
\usepackage{subfigure}  
\usepackage{hyperref}   

\usepackage{braket}

\newcommand{\ba}{\begin{eqnarray}}
\newcommand{\ea}{\end{eqnarray}}
\newcommand{\be}{\begin{equation}}
\newcommand{\ee}{\end{equation}}

\begin{document}

\title{Kuramoto model with run-and-tumble dynamics} 

\author{Derek Frydel}
\affiliation{Department of Chemistry, Universidad Técnica Federico Santa María, Campus San Joaquin, Santiago, Chile}

\date{\today}

\begin{abstract}
This work considers an extension of the Kuramoto model with run-and-tumble dynamics --- a type of self-propelled motion.  
The difference between the extended and the original model is that in the extended version angular velocity of individual 
particles is no longer fixed but can change sporadically with a new velocity drawn from 
a distribution $g(\omega)$.   
Because the Kuramoto model undergoes phase transition, it offers a simple case study for investigating
phase transition for a system with self-propelled particles.  
\end{abstract}

\pacs{
}

\maketitle

\section{Introduction}

This work considers an extension of the Kuramoto model that incorporates self-propelled dynamics.  
Originally conceived as a model of synchronization \cite{Kuramoto75,Kuramoto84,Strogatz91,Strogatz00,Crawford94,Spigler05}, 
the Kuramoto model consists of particles moving on a circle with   
angular velocities distributed according to $g(\omega)$.  Even if particles move  
incoherently, due to coupling interactions, the system transitions 
to a coherent state, in which a fraction of particles locks onto the same angular velocity.  The 
critical value of a coupling constant where the transition transpires can be calculated exactly 
using linear analysis \cite{Strogatz91}.  

Self-propelled dynamics is incorporated into the Kuramoto model 
by introducing a linear "reaction" term into the Fokker-Planck equation (FP).  Without the reaction
term, an angular velocity of a given particle does not change in time.  The source of disorder comes
from the fact that different particles have different velocities in accordance with some distribution 
$g(\omega)$.  This creates a dynamic quenched disorder \cite{Frydel21}.  
With the reaction term, angular velocity of individual particles changes in 
the course of time by sampling the distribution $g(\omega)$.  
At the microscopic level, this means that individual particles change velocity at 
intervals drawn from a Poisson distribution.  Then a new velocity is drawn from the 
distribution $g(\omega)$.

The proposed extension of the Kuramoto model is closely related to a model of self-propelled particles known 
as the run-and-tumble particles (RTP) \cite{Berg83}. In this model, particles are subject to a drift of constant velocity 
but changing orientation.  In 1D, where only two orientations are possible, the model can be solved exactly for 
particles confined between two walls  
\cite{Berg83,Schnitzer93,Weiss02,Cates08,Cates09,Angelani17,Dhar18,Dhar19,Razin20,Basu20,Basu20b}. 
Particles in this model are ideal, that is, non-interacting, and so there can be no phase transition yet, despite its 
simplicity, the model accurately captures many features of self-propelled motion.  
For example, it captures the deposition of particles in a steady-state near the confining walls
\cite{Ebeling00,Hagan14,Brady16,rudi18,Orlandini18}, 
a feature that is not accounted for by a Boltzmann distribution \cite{Frydel21}.  

The Kuramoto model is also a 1D model.  
However, there are some important differences between such a model and the standard RTP in 1D. 
Unlike the RTP-1D model, spacial confinement in the Kuramoto system is not imposed by rigid walls but
is the result of periodic boundary conditions.  Also, unlike in the RTP model in 1D, particles in the Kuramoto model are not 
ideal but interact with each other via a soft attractive pair potential of the form $u_{ij} \propto -\frac{K}{N}\cos(\theta_j-\theta_i)$.  
Because strength of interactions is rescaled by the number of particles $N$ \cite{Yan11}, the 
system is prevented from thermodynamic collapse in the limit $N\to\infty$.  Such 
collapse is common for particles with attractive interactions but no hard-core or some other sort of divergent 
repulsion \cite{Ruelle66,Ruelle99,Frydel18a,Frydel20a}.  Instead of collapsing, the system undergoes phase 
transition from a uniform to heterogeneous distribution (from a coherent to incoherent state).  
The phase transition in the mean-field limit can be determined exactly.


This paper is organized as follows.  In Sec. (\ref{sec:1}) we introduce the Kuramoto model 
and the extension to run-and-tumble particles.  In Sec. (\ref{sec:2}) we consider a simple
situation with both coupling parameters, $K$ and $\alpha$, set to zero.  Then in Sec. (\ref{sec:3}) 
we consider the situation $\alpha>0$ and $K=0$.  
In Sec. (\ref{sec:4}) we consider the complete model and analyze it using linear theory and then present
numerical results.  Finally, in Sec. (\ref{sec:5}) we conclude the work.  


\section{The Kuramoto model}
\label{sec:1}

The Kuramoto model is a model of synchronization stripped to a mathematical minimum 
\cite{Kuramoto75,Kuramoto84,Strogatz91,Strogatz00,Crawford94,Spigler05}. 
It consists of a population of $N$ coupled oscillators with phase $\theta_i$ and frequency
$\omega_i$ distributed with a given probability $g(\omega)$.  
In addition, particles interact with each other so that individual frequencies are correlated.  
Dynamics of the model is governed by the following equation 
\be
\frac{d\theta_i(t)}{dt} = \omega_i +  \frac{K}{N} \sum_{j=1}^N \sin(\theta_j-\theta_i),
\ee
where 
$K$ is the  coupling strength.  In the limit $N\to \infty$, the system is described exactly by the mean-field 
approximation.   As different particles have different velocities $\omega_i$ 
distributed according to $g(\omega)$, the system possesses a dynamic quenched disorder, 
where $\omega_i$is the disorder variable.

A common extension of the model is to include Gaussian white noise \cite{Sakaguchi88}, 
\be
\frac{d\theta_i(t)}{dt} = \omega_i +  \xi_i(t) + \frac{K}{N} \sum_{j=1}^N \sin(\theta_j-\theta_i), 
\label{eq:LS}
\ee
such that 
\be
\langle \xi_i(t)\rangle = 0,~~~~ \langle \xi_i(t)\xi_j(t')\rangle = 2D\delta_{ij} \delta(t-t'),
\label{eq:noise}
\ee
and where $D$ is the diffusion constant.

One may construe Eq. (\ref{eq:LS}) as a Langevin 
equation for particles moving on a circle with angular velocity $\omega_i$ 
and interacting with each other via a pair potential $\beta u_{ij} = -\frac{K}{D}\cos(\theta_j-\theta_i)$, 
where $\beta = 1/k_BT$.  An advantage of this particular functional form of interactions is that it 
does not change when a given particle interacts with one, two, or a distribution of particles $n(\theta)$, for example, 
$\int d\theta'\, n(\theta) \cos(\theta-\theta_i) \propto \cos(\psi-\theta_i)$, where $\psi$ is the center
of mass of $n(\theta)$.  This feature makes the model conducive to mean-field treatment.  

Within the Fokker-Planck formulation, the system can be represented as  
\ba
\frac{\partial \rho}{\partial t} &=& D\frac{\partial^2 \rho}{\partial \theta^2} 
- \frac{\partial}{\partial \theta} \bigg[\rho \bigg(\omega+ K r \sin(\psi-\theta)\bigg) \bigg], 
\label{eq:FP0}
\ea
where $\rho\equiv \rho(\theta,\omega,t)$ is the normalized distribution and 
\be
r \sin(\psi-\theta) = \int_0^{2\pi} d\theta'\, \sin(\theta'-\theta) 
\int_{-\infty}^{\infty} d\omega \, g(\omega) \rho(\theta',\omega,t).  
\label{eq:r0}
\ee
The quantity $r$ is considered as an amplitude of an order parameter and $\psi$ as its phase.  
Obviously, $r$ is not known {\it a priori} but depends on a density.  
Eqs. (\ref{eq:FP0},\ref{eq:r0}) represent a set of self-consistent relations.  

The point where $r$ first becomes non-zero corresponds to a phase transition.  The transition is 
from an incoherent state, represented by a uniform 
distribution $\rho=1/2\pi$ and $r=0$, to a coherent state, represented by a heterogenous 
distribution and $r>0$.  The critical value of $K$ where this occurs is determined from the following
relation (assuming that $g(\omega)$ is unimodal and with even symmetry)
\cite{Sakaguchi88,Strogatz91,Strogatz00,Crawford94,Spigler05} 
\be
K_c  
= 2 \bigg[\int_{-\infty}^{\infty} d\omega\,  g(\omega)\frac{ D}{D^2+\omega^2} \bigg]^{-1}.  
\label{eq:Kc}
\ee


The distribution of frequencies (or angular velocities) $g(\omega)$ introduces dynamic 
quenched disorder, while the Gaussian white noise introduces Brownian fluctuations.  For the case without 
quenched disorder, represented by the distribution $g(\omega)=\delta(\omega)$, the critical coupling obtained
from Eq. (\ref{eq:Kc}) is $K_c=2D$.  On the other hand, for a system with quenched disorder but without 
Gaussian noise, $D = 0$, Eq. (\ref{eq:Kc}) evaluates to $K_c=\frac{2}{\pi g(0)}$.

One of the goals of this article is the derivation of an analogous relation to that in Eq. (\ref{eq:Kc}) for an extended 
Kuramoto model that includes run-and-tumble
type of dynamics developed and analyzed in this work.  The inclusion of run-and-tumble  
dynamics, which permits particles to sporadically change their angular velocity $\omega_i$ with a rate $\alpha$, 
can be considered as a third source of disorder, in addition to the quenched disorder and the 
Gaussian noise.

\subsection{Kuramoto model for self-propelled particles}
\label{sec:kuramoto2}

As stated above, 
in this work we consider an extension of the Kuramoto model that incorporates self-propelled motion ---
or more specifically, run-and-tumble type of dynamics.  While in the original model, 
governed by the FP equation (\ref{eq:FP0}), an angular velocity of an individual particle $\omega_i$ is fixed, 
in the extended version, individual angular velocities are allowed to evolve in time.  As a consequence, 
every particle can sample velocities of the distribution $g(\omega)$. 

On the microscopic level this means that a particle changes velocity at time intervals drawn from the 
Poisson distribution $\propto e^{-\alpha t}$, where $\alpha$ is the frequency at which this event takes 
place.  A new velocity is then randomly drawn from the distribution $g(\omega)$.  This type of dynamics
corresponds to the run-and-tumble type of motion, one of the standard models of self-propelled particles 
\cite{Berg83,Schnitzer93,Weiss02,Cates08,Cates09,Angelani17,Dhar18,Dhar19,Razin20,Basu20,Basu20b}.  
What might be different in our version of the run-and-tumble dynamics, compared to more conventional 
ways it is implemented, is that the distribution of angular velocities $g(\omega)$ is arbitrary.  

We note that the run-and-tumble dynamics is linked to the distribution $g(\omega)$ and, thereby, to quenched 
disorder of a system \cite{Frydel21}.  If a distribution is $g(\omega)=\delta(\omega)$, therefore, there is no
quenched disorder, then the run-and-tumble dynamics is no longer possible, no matter what value of the 
parameter $\alpha$.  Run-and-tumble dynamics is about how fast a single particle can sample a system's 
quenched disorder.

The extension of the Kuramoto model  
just described is most conveniently incorporated within the Fokker-Planck formulation.  
This is done by including a linear "reaction" term to the FP equation in (\ref{eq:FP0}), resulting in 
\ba
\frac{\partial \rho}{\partial t} &=& D\frac{\partial^2 \rho}{\partial \theta^2} 
- \frac{\partial}{\partial \theta} \bigg[\rho \bigg(\omega+ K r \sin(\psi-\theta)\bigg) \bigg]  
+ \alpha ( \bar \rho - \rho),\nonumber\\
\label{eq:FPR}
\ea
where to simplify expressions, we introduce the average density defined as 
\be
\bar\rho(\theta,t) =  \int_{-\infty}^{\infty} d\omega \, g(\omega) \rho(\theta',\omega,t).  
\label{eq:barn}
\ee
If particles with different $\omega$ are interpreted as different species, then the "reaction" 
term can be viewed as a process that converts one type of particle into another.  

It is impossible, based on a simple inspection of Eq. (\ref{eq:FPR}), to predict 
how the parameter $\alpha$ should modify Eq. (\ref{eq:Kc}).  Both $K$ and $\alpha$ act as coupling 
parameters, that is, they both couple different distributions $\rho(\theta,\omega,t)$ for different $\omega$.  
This produces an expectation that by enhancing coupling, $\alpha$ should lower the critical point $K_c$.  
On the other hand, the rate $\alpha$ that controls the frequency with which a particle changes its
angular velocity, could be regarded as a diffusion enhancing contribution, in which role it should increase 
the critical point $K_c$.  



The motivation to consider such an extension of the Kuramoto model is to gain deeper and 
more fundamental understanding of self-propelled motion by considering it in different settings.  
The Kuramoto model, in particular, provides an interesting case study due to occurrence of a phase 
transition.  In consequence, it offers a simple setting for studying critical phenomenon with 
participation of self-propelled motion.  

\section{The case $K=0$ and $\alpha=0$}
\label{sec:2}

We start by considering a simple scenario:  the Kuramoto model with both coupling parameters set to zero, 
$K=\alpha=0$.   
Eq. (\ref{eq:FPR}) in this situation reduces to a diffusion-convection equation
\be
\frac{\partial \rho}{\partial t} = D\frac{\partial^2 \rho}{\partial \theta^2} - \omega \frac{\partial\rho}{\partial \theta}.  
\label{eq:FP00}
\ee
For the initial distribution 
\be
\rho(\theta,\omega,0) = \delta(\theta),
\label{eq:rho0}
\ee
that is, for all particles initially placed at $\theta_i=0$, 
the solution is a propagating Gaussian distribution 
\be
\rho(\theta,\omega,t) =  \frac{e^{-(\theta - \omega t)^2/4Dt}}{\sqrt{4\pi Dt}}.  
\label{eq:rho_dc}
\ee 

Note that the above solution ignores periodic boundary conditions, that is, $\rho(\theta+2\pi,\omega,t)\neq \rho(\theta,\omega,t)$.  
As we are not interested 
in the distribution $\rho(\theta,t)$ {\it per se} but quantities derived from it, the above expression is sufficient to
our purposes.

The quantity that is of interest is $r$ defined earlier in (\ref{eq:r0}).  
It not only measures the extent of interactions in both Eq. (\ref{eq:FP0}) and Eq. (\ref{eq:FPR})  
but also plays the role of the order parameter of a phase transition.  
Below, we define $r$ a little differently from the definition in (\ref{eq:r0}) 
\be
r(t)e^{-i\psi}  = \int_{-\infty}^{\infty} d\omega \, g(\omega) \int_{-\pi}^{\pi} d\theta \, \rho(\theta,\omega,t) e^{-i\theta}. 
\label{eq:rt0}
\ee
But if we want to use the solution in (\ref{eq:rho_dc}), without periodic boundary conditions, we 
have to modify the above integral as 
\be
r(t) e^{-i\psi}  = \int_{-\infty}^{\infty} d\omega \, g(\omega) \int_{-\infty}^{\infty} d\theta \, \rho(\theta,\omega,t) e^{-i\theta}.  
\label{eq:rt0a}
\ee
As periodic boundary conditions are implicit in $e^{-i\theta}$, we are justified to 
ignore the periodicity in $\rho$.  Substituting the solution in (\ref{eq:rho_dc}) 
into a modified definition for $r$ in Eq. (\ref{eq:rt0a}) yields 
\be
r(t)
=  e^{-D t}  \int_{-\infty}^{\infty} d\omega \, g(\omega) e^{-i\omega t}.
\label{eq:rt1}
\ee
For $r$ to be real valued, $g(\omega)$ ought to have an even symmetry.  


The above result tells us how $r(t)$ evolves in time.  At time $t=0$, for the initial distribution in (\ref{eq:rho0}), $r=1$.  
Without coupling between particles, there can be no phase transition and at long times
$r(t)\to 0$.  The above expression distinguishes between two mechanisms of relaxation:   
the collisional relaxation that produces exponential decay $e^{-D t}$, and the collisionless relaxation
that involves simple mixing as a result of quenched disorder, arising as a result of 
distribution of angular velocities $g(\omega)$.  
The collisionless mechanism depends on particular functional form of $g(\omega)$ \cite{Strogatz92}.

\subsubsection{concrete examples}
\label{sec:CC1}

For a system without quenched disorder, represented 
by a singular distribution $g(\omega)= \delta(\omega)$, Eq. (\ref{eq:rt1}) evaluates to 
\be
r(t) =  e^{-D t}.  
\label{eq:r_delta}
\ee
Here, the only mechanism of relaxation is collisional dissipation producing exponential decay.  

Next, we consider a Lorentz distribution, 
$g(\omega)=\frac{1}{\pi} \frac{\omega_0}{ \omega_0^2 + \omega^2}$.  In this case 
Eq. (\ref{eq:rt1}) evaluates to 
\be
r(t) = e^{-(D+\omega_0) t}. 
\label{eq:rL}
\ee
For this type of quenched disorder, the relaxation due to collisionless mechanism is exponential, like
that for collisional mechanism.  The two processes are, therefore, compatible, and we can think of 
$D+\omega_0$ as an effective diffusion.  

For a Gaussian distribution, $g(\omega) = \frac{e^{-\omega^2/2\omega_0^2}}{\sqrt{2\pi \omega_0^2}}$, 
Eq. (\ref{eq:rt1}) evaluates to 
\be
r(t) =  e^{-D t} e^{-\omega_0^2 t^2/2}.  
\label{eq:r_gauss}
\ee
Even though the collisionless and collisional relaxation have different functional form, 
both processes are fast.

A uniform distribution, $g(\omega) = \frac{1}{2\omega_0}$ defined on the interval $-\omega_0\le \omega\le \omega_0$, 
is somewhat different from the two cases above.  Eq. (\ref{eq:rt1}) for this type of quenched disorder evaluates to 
\be
r(t) = e^{-D t} \frac{ |\sin \omega_0 t|}{\omega_0 t}.  
\label{eq:r_uni}
\ee
The collisionless relaxation in this case has a more interesting behavior;  its decay is algebraic and it exhibits 
oscillations.  

Another distribution frequently considered in the context of the Kuramoto model is a discrete binodal distribution 
$g(\omega) = \frac{1}{2}\delta(\omega-\omega_0) +  \frac{1}{2}\delta(\omega+\omega_0)$
\cite{Okuda91,Bonilla92,Bonilla98}.  Eq. (\ref{eq:rt1}) in this case yields 
\be
r(t) = e^{-D t} |\cos (\omega_0 t)|.  
\label{eq:r_bi}
\ee
For this distribution, the collisionless relaxation mechanism due to mixing 
does not exist.  It would seem that a collisionless mechanism requires a continuous distribution $g(\omega)$.

\section{The case $\alpha>0$ and $K=0$}
\label{sec:3}

Next, we consider the Kuramoto model with run-and-tumble type of motion but without other type of interactions, $K=0$.  
The FP equation describing this situation is 
\be
\frac{\partial \rho}{\partial t} = D\frac{\partial^2 \rho}{\partial \theta^2} 
- \omega \frac{\partial\rho}{\partial \theta} + \alpha ( \bar \rho - \rho).  
\label{eq:FP000}
\ee
As we are interested in the behavior of the order parameter $r$, we will transform the above equation into 
an equivalent relation but in terms of $r$.  

We proceed by operating on both sides of Eq. (\ref{eq:FP0}) with $\int_{-\pi}^{\pi} d\theta\, e^{-i\theta}$.  
This amounts to Fourier transforming the FP equation with respect to the wavenumber $k=1$.   The transformed 
equation is 
\be
\frac{\partial c_1}{\partial t} = -\big(D + \alpha + i\omega \big)c_{1}  + \alpha r(t) e^{-i\psi}
\label{eq:dc_1}
\ee
where 
\be
c_1(\omega,t) = \int_{-\pi}^{\pi} d\theta\, \rho(\theta,\omega,t) e^{-i\theta},
\label{eq:c_1}
\ee
The last term in (\ref{eq:dc_1}) comes from the definition 
$
re^{-i\psi}  = \int_{-\infty}^{\infty} d\omega \, g(\omega) c_1(\omega,t).
$
If regarded as a first order inhomogeneous equation, which is possible if we ignore the fact that 
$r$ is a functional of $c_1$, then the solution to Eq. (\ref{eq:dc_1}) 
for the initial distribution in (\ref{eq:rho0}), can be represented as 
\ba
c_1(\omega,t)  &=& e^{-i\theta_0} e^{-(D + \alpha + i\omega) t} \nonumber\\ 
&+& \alpha e^{-i\psi} \int_0^{t} dt'\, e^{-(D + \alpha + i\omega) (t-t')} r(t').
\label{eq:dc_1b}
\ea
Finally, by operating on the above equation with $\int_{-\infty}^{\infty} d\omega\,g(\omega)$ 
we get 
\be
r(t) = R(t) + \alpha\int_0^{t} dt'\, r(t') R(t-t').  
\label{eq:rt2}
\ee
The result is a convolution equation where the kernel $R$ is given by 
\be
R(t)  =  e^{-(D+\alpha) t} \int_{-\infty}^{\infty} d\omega \, g(\omega) e^{-i\omega t}.  
\label{eq:r00}
\ee
For $\alpha=0$, Eq. (\ref{eq:rt2}) recovers the result in (\ref{eq:rt1}).  For $\alpha>0$, the 
evolution of $r(t)$ involves a kernel that is expected to slow down the relaxation of $r$.  


The behavior of $r$, as determined by Eq. (\ref{eq:rt2}), depends on a particular type of 
quenched disorder, that is, a particular distribution $g(\omega)$.  For a singular, Lorentz, discrete 
bimodal distribution, the equation can be solved exactly.  For a uniform and Gaussian distributions 
exact expression doesn't seem possible, or at least it is not straightforward.  
For those cases, we focus on an analysis of an asymptotic behavior at long times.  

\subsubsection{Laplace analysis}
We start by pointing out that Eq. (\ref{eq:rt2}) represents the Volterra integral equation of the second kind.  
A common method of analyzing this type of equation is by using the Laplace transform techniques.  This is 
the approach that we are going to take.

We start by recalling the a Laplace transformed function $f(t)$ is defined as 
$\hat f(s) = \int_0^{\infty} dt\, e^{-st} f(t)$. 
Taking the Laplace transform of Eq. (\ref{eq:rt2}) yields 
$$
\hat r(s) = \frac{\hat R(s)}{1 - \alpha  \hat R(s)}, 
$$
where 
$$
 \hat R(s) = \int_{-\infty}^{\infty} d\omega \, \frac{g(\omega)}{s + D + \alpha + i\omega}.  
$$
To obtain an expression of $r$ in real time, we use the inverse Laplace transform \cite{Frydel19} 
leading to 
\be
r(t) = \frac{1}{2\pi i} \lim_{T\to\infty} \int_{a-iT}^{a+iT} ds\,  \frac{ \hat R(s) e^{s t} }{1-\alpha \hat R(s)}.  
\label{eq:rt_il}
\ee

The above expression has an advantage that it can be analyzed 
using the residue theorem that boils down to identification of the poles $s_p$ and $r$ can be represented in 
terms of residues at those poles as 
$$
r(t) = \sum_{s_p} \text{Res}\bigg[\frac{ \hat R(s) e^{s t} }{1-\alpha \hat R(s)}\bigg].  
$$
To make the integral in (\ref{eq:rt_il}) more intelligible, 
we explicitly represent $\hat R(s)$ of the numerator, yielding 
\be
r(t) = \frac{1}{2\pi i} \int_{\gamma-iT}^{\gamma+iT} 
 \frac{ds\,  e^{s t}}{1-\alpha \hat R(s)}  \int_{-\infty}^{\infty} \frac{d\omega \, g(\omega)}{s + D + \alpha + i\omega}.
\label{eq:rt_il2}
\ee



The above expression allows us to distinguish two types of poles.  The poles of the second fraction, 
\be
s_c = -D-\alpha-i\omega,
\label{eq:sc}
\ee
are continuous by virtue of the integral over $\omega$.  On a complex plane, those poles are represented by a line 
parallel to an imaginary axis and offset to the left by $-D-\alpha$.  

Discrete poles of the second fractional term, on the other hand, satisfy the relation 
$1 = \alpha \hat R(s_d)$, which if written explicitly leads to the following relation 
\be
1 = \alpha  \int_{-\infty}^{\infty} d\omega \, \frac{g(\omega)}{s_d + D + \alpha + i\omega}.  
\label{eq:sp}
\ee

Note that $\alpha$ appears in two different places in Eq. (\ref{eq:rt_il2}) -- the fact we have already alluded to before.  
On the one hand, $\alpha$ enhances the diffusion constant $D$.  On the other hand, it appears separately from 
$D$ where it plays the role of a coupling parameter.  The coupling function of $\alpha$ is captured by discrete 
poles $s_d$.  Consequently, we restrict our analysis to $s_d$.



Prior to considering different concrete cases, we indicate that if $g(\omega)$ has even symmetry and is 
unimodal then there can be at most one pole $s_d$ whose value is real \cite{Strogatz91,Strogatz00}.  
For the case of a discrete bimodal $g(\omega)$, there are two poles $s_d$ that are not restricted 
to a real value \cite{Bonilla92,Bonilla98}.

\subsubsection{concrete examples}

We start with a singular distribution $g(\omega)=\delta(\omega)$ --- a system without quenched 
disorder.  Eq. (\ref{eq:rt2}) in this case is solved exactly where, unsurprisingly, it recovers the result 
in (\ref{eq:r_delta}) for a system for $\alpha=0$, 
\be
r(t) = e^{- Dt}.  
\label{eq:rs}
\ee
Without quenched disorder there can be no run-and-tumbling dynamics.  

For a Lorentz distribution, Eq. (\ref{eq:rt2}) is solved exactly, leading to 
$$
r(t) = e^{- Dt}e^{-\omega_0 t}.  
$$
This is the same result as that in (\ref{eq:rL}) for $\alpha=0$, implying that the run-and-tumble dynamics 
does not alter the evolution of $r(t)$.  This is different from the case $g(\omega)=\delta(\omega)$, where 
the run-and-tumble dynamic simply does not exist.  If we focused on 
a single particle trajectory (for a system with a Lorentz distribution), we would find that trajectories for 
different $\alpha$ are very different.  Yet when considering collectively, by looking at the evolution of $r$, 
we detect no change.  This unusual result, rather than being general is a feature of a Lorentz 
distribution.  

To see this, we consider next a uniform distribution on the interval $-\omega_0 \le \omega\le \omega_0$.  
Because Eq. (\ref{eq:rt2}) cannot be solved exactly, we analyze an asymptotic 
behavior, $r(t) \approx e^{s_d t}$, determined by a discrete pole $s_d$.  
From the relation (\ref{eq:sp}) we get 
\be
s_d = -(D+\alpha) + \omega_0 \cot(\omega_0/\alpha), 
\label{eq:spu}
\ee
implying the following long time relaxation $r \propto e^{-(D+\alpha)t} e^{ t  \omega_0  \cot( \omega_0 /\alpha)}$.  
Compared to evolution of $r$ in Eq. (\ref{eq:r_uni}) for $\alpha=0$, we see that the run-and-tumble dynamics 
modifies the functional form from algebraic oscillatory to exponential monotonic.  

Change of a functional form implies a discontinuity that occurs 
at some specific value of $\alpha$, which we refer to as a point of crossover, $\alpha_{cross}$.  
A crossover can be determined from Eq. (\ref{eq:sp}) by noting that for $g(\omega)$ that is 
unimodal and with even symmetry $s_d$ is real valued.  This permits us to rewrite Eq. (\ref{eq:sp}) as 
\be
1 = \alpha  \int_{-\infty}^{\infty} d\omega \, g(\omega)\frac{s_d + D + \alpha}{(s_d + D + \alpha)^2 + \omega^2}.  
\label{eq:sp3}
\ee
The above integral can be interpreted as an overlap integral between two normalized distributions, 
$g(\omega)$ and the Lorentz distribution.  Since $s_d \ge -D - \alpha$ (if $s_d < -D - \alpha$, the integral term  
becomes negative and equality cannot be satisfied), we may assume that the crossover occurs at the border 
value $s_d = -D - \alpha$.  
In such a case, the Lorentz distribution transforms into a delta function, leading to $1 = \pi \alpha_{cross} g(0)$.  
Consequently, we may write 
\be
\alpha_{cross} = \frac{1}{\pi g(0)}.  
\label{eq:alpha_cross}
\ee


We will next establish that the coupling due to a finite $\alpha$ cannot produce phase transition, 
that is, there is no finite value of $\alpha$ that yields $s_d=0$.  
If we take the limit $\alpha\to\infty$, Eq. (\ref{eq:sp3}) reduces to 
\be
1 \approx \frac{\alpha}{s_d + D + \alpha}.  
\ee
The limiting value of $s_d$ is $s_d = -D$ which is approached from below.  This means that $s_d=0$ can only 
occur if $D=0$ and $\alpha\to \infty$.  We can, therefore, exclude any phase transition.  
As $s_d=-D$ corresponds to a system without quenched disorder, see Eq. (\ref{eq:rs}), this means 
that in the limit $\alpha\to \infty$ quenched disorder is completely eliminated.  

\begin{figure}[h] 
 \begin{center}
 \begin{tabular}{rrrr}
\includegraphics[height=0.21\textwidth,width=0.25\textwidth]{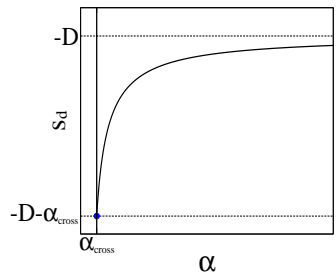} &&
 \end{tabular}
 \end{center} 
\caption{Discrete pole $s_d$ as a function of $\alpha$ for a uniform distribution $g(\omega)$.  
The results correspond to Eq. (\ref{eq:spu}) for parameters $D=1$ and $\omega_0=1$. }
\label{fig:fig1} 
\end{figure}

Separating terms in Eq. (\ref{eq:spu}) that depend on $\alpha$, 
we may find that the contributions of a run-and-tumble motion vanish in the limit $\alpha\to \infty$, 
that is, $\omega_0 \cot(\omega_0/\alpha) - \alpha\to 0$.  
In Fig. (\ref{fig:fig1}) we plot Eq. (\ref{eq:spu}) as a function of $\alpha$.  
\graphicspath{{figures/}}

The fact that large $\alpha$ eliminates a quenched disorder is not 
surprising to anyone familiar with propelled particles wherein the limit $\alpha\to\infty$ is considered
as an equilibrium state where stationary distributions recover Boltzmann functional 
form \cite{Razin20,Frydel21}.

\begin{figure}[h] 
 \begin{center}
 \begin{tabular}{rrrr}
\includegraphics[height=0.21\textwidth,width=0.25\textwidth]{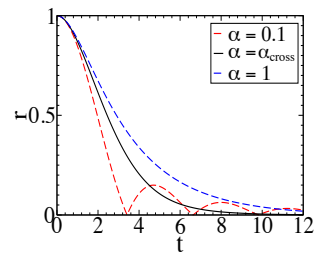} &&
 \end{tabular}
 \end{center} 
\caption{Evolution of $r(t)$ for a uniform $g(\omega)$.  The system parameters are $D=0$ and $\omega_0=1$.
The results are from dynamic simulation for $N=10^6$ particles and time interval
$dt=0.001$ with the initial configuration corresponding to all particles placed at $\theta_i=0$.}
\label{fig:fig2} 
\end{figure}

In Fig, (\ref{fig:fig2}) we plot the evolution of $r(t)$ for a uniform distribution for 
three different values of $\alpha$: below, above, and at the crossover 
value of $\alpha$.  For a uniform distribution, $\alpha_{cross} = 2\omega_0/\pi$.
\graphicspath{{figures/}}

For a Gaussian distribution $g(\omega)$ we get a similar behavior to that for a uniform distribution.
The crossover point where the functional form in (\ref{eq:r_gauss}) changes to an exponential 
decay is obtained using Eq. (\ref{eq:alpha_cross}).  Then as $\alpha$ approaches infinity, quenched
disorder is eliminated.  

As a last example, we consider a discrete bimodal distribution already introduced at the end of Sec. (\ref{sec:CC1}).  
Eq. (\ref{eq:rt2}) for this distribution is solved exactly and the result is 
$$
r(t) = e^{-(D + \frac{\alpha}{2})t} \bigg|\cos(\omega_e t) + \frac{1}{2}\frac{\alpha}{\omega_e}  \sin(\omega_e t) \bigg|, 
$$
where 
$$
\omega_e = \omega_0 \sqrt{1 - \bigg(\frac{\alpha}{2\omega_0}\bigg)^2 }. 
$$
This is a solution for a damped oscillator.  
The parameter $\alpha$ affects the change from an underdamped to overdamped dynamics at $\alpha = 2\omega_0$.  

\section{Finite $K$ and $\alpha$}
\label{sec:4}

The Kuramoto model with both coupling parameters set to finite value, $\alpha>0$ and $K>0$, 
is governed by the following FP equation 
\be
\frac{\partial \rho}{\partial t} = D\frac{\partial^2 \rho}{\partial \theta^2} 
- \omega \frac{\partial\rho}{\partial \theta} + \alpha ( \bar \rho - \rho)
- K r\frac{\partial \rho \sin(\psi-\theta)}{\partial \theta}.
\label{eq:FPK}
\ee
To apply a linear analysis of the previous section, the last term in the equation is linearized 
by representing the density as $\rho = \frac{1}{2\pi} + \delta\rho$, where $\delta \rho$ is 
the deviation from a uniform density.  The linearized last term then becomes  
$-K r\frac{\partial \rho \sin(\psi-\theta)}{\partial \theta} \approx \frac{K}{2\pi} r\cos(\psi-\theta)$, and the 
corresponding linearized FP equation is 
\be
\frac{\partial \rho}{\partial t} = D\frac{\partial^2 \rho}{\partial \theta^2} 
- \omega \frac{\partial\rho}{\partial \theta} + \alpha ( \bar \rho - \rho)
+ \frac{K}{2\pi} r\cos(\psi-\theta).  
\label{eq:FPK}
\ee

Following the steps in Eq. (\ref{eq:dc_1}) and Eq. (\ref{eq:dc_1b}) we arrive at an analogous result to that 
in (\ref{eq:rt2})
\be
r(t) = R(t) + \bigg(\frac{K}{2}+\alpha\bigg) \int_0^{t} dt'\, r(t') R(t-t'), 
\label{eq:rt3}
\ee
with $R(t)$ is defined in (\ref{eq:r0}).  The equation is next transformed using the Laplace transform techniques into 
\ba
r(t) = \frac{1}{2\pi i} 
\int_{\gamma-iT}^{\gamma+iT} ds\,  
\frac{e^{s t}  \hat R(s)}{1-(\alpha+K/2) \hat R(s)}, 
\label{eq:rt_il2}
\ea
where the discrete pole is obtained from the following relation 
\be
1 = \bigg(\frac{K}{2}+\alpha\bigg) \int_{-\infty}^{\infty} d\omega\,   \frac{g(\omega)}{s_d + D + \alpha + i\omega}.  
\label{eq:sp2}
\ee
The above relation can subsequently be used to obtain a critical value of $K$ where the incoherent solution 
becomes unstable.  Assuming that $g(\omega)$ is unimodal with even symmetry, this occurs when $s_d=0$ leading to 
\be
1 = \bigg(\frac{K_c}{2}+\alpha\bigg) \int_{-\infty}^{\infty} d\omega\,   g(\omega)\frac{D + \alpha}{(D + \alpha)^2 + \omega^2}.  
\label{eq:sp2a}
\ee

The above relation is a central result of this article.  It is analogous to a similar relation for the Kuramoto model 
without self-consistent dynamics, see Eq. (\ref{eq:Kc}).  It shows how the onset of self-propelled motion modifies 
a critical point.  The parameter $\alpha$ appears in two places, suggesting 
two different roles.  On the one hand, it functions as an enhancement of diffusion $D$.  
On the other hand, it enhances the coupling parameter $K$.   The two roles work in opposite directions.  
Enhanced dissipation is expected to increase the critical value $K_c$ (increased
dissipation means stronger coupling is required to bring about the coherent state), while enhanced 
coupling is expected to reduce the critical value $K_c$.  

\subsubsection{concrete examples}

In the case of a Lorentz distribution, $K_c$ does not depend on $\alpha$, 
and Eq. (\ref{eq:sp2a}) in this case leads to
$$
K_c = 2(D+\omega_0).  
$$
Even though dynamics of individual particles is a function of $\alpha$, when it comes to collective 
dynamics, in the case of a Lorentz distribution, no change can be detected.

\begin{figure}[h] 
 \begin{center}
 \begin{tabular}{rrrr}
\includegraphics[height=0.21\textwidth,width=0.25\textwidth]{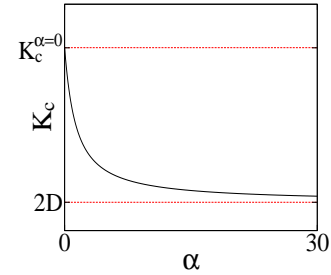} &&
 \end{tabular}
 \end{center} 
\caption{$K_c$ as a function of $\alpha$ as given in (\ref{eq:Kc_uni}) for a uniform $g(\omega)$.  The parameters are $D=1$ 
and $\omega_0=1$.}
\label{fig:fig3} 
\end{figure}

For a uniform distribution $g(\omega)$, the relation in (\ref{eq:sp2a}) yields 
\be
K_c  = \frac{2\omega_0}{\arctan\big(\frac{\omega_0}{D+\alpha} \big)} - 2\alpha.  
\label{eq:Kc_uni}
\ee
As the first term increases with increasing $\alpha$, the second term produces an opposite trend  
In the first term, $\alpha$ enhances diffusion, and in the second term it 
enhances coupling between particles.  The net behavior is seen in Fig. (\ref{fig:fig3}) where we plot $K_c$ in 
Eq. (\ref{eq:Kc_uni}) as a function of $\alpha$.  
\graphicspath{{figures/}}
The plot shows a monotonically decreasing $K_c$, indicating that the contribution of $\alpha$ to coupling is 
a dominant factor.  For large $\alpha$, the dependence of $K_c$ on $\alpha$ is
$$
\lim_{\alpha\to \infty} K_c \approx 2D + \frac{2\omega_0^2}{3} \frac{1}{\alpha}, 
$$
where the limiting value of $K_c$ is $2D$.  As discussed in Sec. (\ref{sec:1}), below Eq. (\ref{eq:Kc}), this 
value corresponds to the system without quenched disorder (when the distribution $g(\omega)$ is singular).  
This once again goes to show that the self-propelled dynamics effectively leads to elimination of 
quenched disorder.  


A similar behavior is observed for a Gaussian distribution $g(\omega)$.  In this case Eq. (\ref{eq:sp2a}) evaluates to 
\be
K_c  = \frac{2\omega_0}{\sqrt{\pi/2}} \frac{e^{-\frac{(D+\alpha)^2}{2\omega_0^2}}}{\text{erfc}\big[\frac{D+\alpha}{\omega_0\sqrt{2}}\big]} - 2\alpha, 
\label{eq:Kc_uni}
\ee
where for large $\alpha$ we have 
$$
\lim_{\alpha\to \infty} K_c \approx 2D +  \frac{2\omega_0^2}{\alpha}, 
$$
indicating a gradual elimination of quenched disorder as $K_c\to 2D$.   



As a final example, we consider a discrete bimodal distribution.  This scenario is more complicated, 
involving multiple bifurcations, full understanding of which requires nonlinear analysis 
\cite{Bonilla92,Crawford94,Bonilla98}.  Here, we limit ourselves to linear analysis and the role 
played by $\alpha$.  

\begin{figure}[h] 
 \begin{center}
 \begin{tabular}{rrrr}
\includegraphics[height=0.21\textwidth,width=0.25\textwidth]{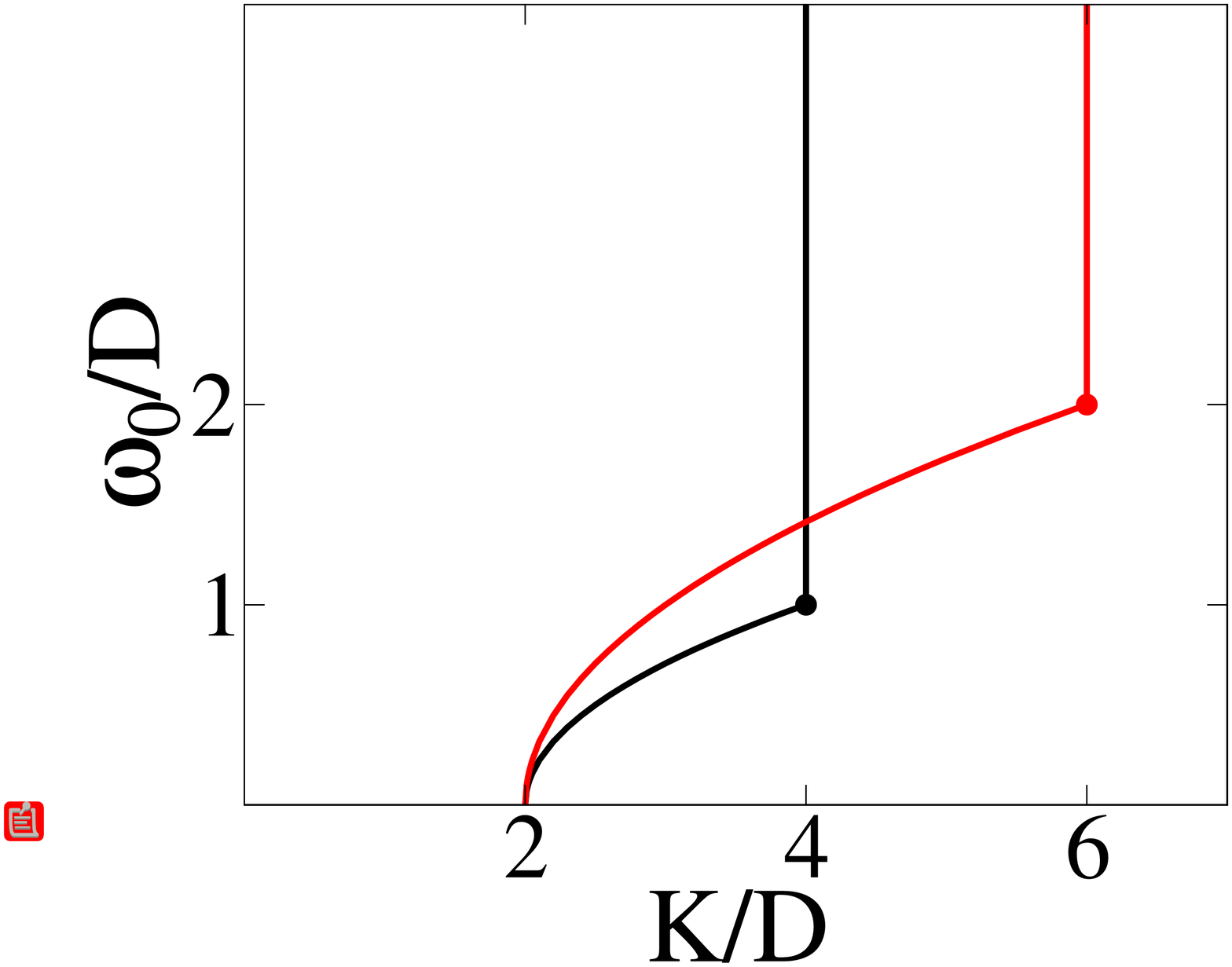} &&
 \end{tabular}
 \end{center} 
\caption{Linear stability diagram for a discrete bimodal distribution $g(\omega)$ in the parameter space $(K/D,\omega_0/D)$.
The incoherent state is linearly stable to the left of the lines.  The black lines are for $\alpha=0$ and red lines for $\alpha=D$.  
The solid circles designate a tricritical point above which the transition corresponds to a Hopf bifurcation represented
by vertical lines in (\ref{eq:KcH}).
A similar diagram for $\alpha=0$ can be found in Fig. 1 of Ref. \cite{Bonilla98}.  }
\label{fig:fig3a} 
\end{figure}

From Eq. (\ref{eq:sp2}) we get 
\be
s_d = -\bigg(D + \frac{\alpha}{2} - \frac{K}{4}\bigg) \pm  i\omega_0 \sqrt { 1 - \bigg(\frac{2\alpha+K}{4\omega_0}\bigg)^2 },
\label{eq:sp_b}
\ee
indicating the existence of two poles.  The result is similar to that found in Eq. (12) of Ref. \cite{Bonilla98} but 
limited to the case $\alpha=0$.  In the regime $4\omega_0 > K + 2\alpha$, the poles are complex.  The incoherent 
state becomes unstable when the real part of $s_d$ vanishes, which corresponds to 
\be
K_c^h = 4 D + 2 \alpha.  
\label{eq:KcH}
\ee
The superscript "h" designates a Hopf bifurcation and involves transformation to a time-periodic behavior of the 
order parameter $r(t)$.  This bifurcation is shifted up as $\alpha$ increases.  
In the regime $4\omega_0 < K + 2\alpha$ where the poles are real, the phase transition occurs when 
the smaller of the two poles becomes zero.  This corresponds to 
\be
K_c = 2 D + \frac{2\omega_0^2}{D+\alpha}.  
\ee
In this case, $K_c$ is shifted down with increasing $\alpha$.  
Phase diagram constructed from (\ref{eq:sp_b}) is shown in Fig. (\ref{fig:fig3a}) for two cases:  $\alpha=0$ and $\alpha=D$.
\graphicspath{{figures/}}

\subsection{The model for $K>K_c$}

\begin{figure}[h] 
 \begin{center}
 \begin{tabular}{rrrr}
\includegraphics[height=0.21\textwidth,width=0.25\textwidth]{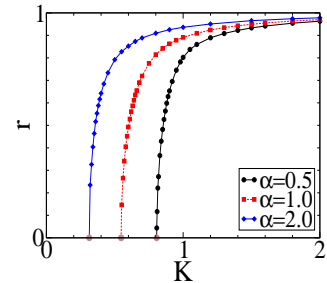} &&
 \end{tabular}
 \end{center} 
\caption{Data points for an average $r$ as a function of $K$ for a uniform $g(\omega)$.  
The data points are obtained from dynamic simulations for $N=10^{5}$ particles and time step $\Delta t=0.01$.   
The system parameters are $\omega_0=1$ and $D=0$.  
The three data points (brown circles) on the $x$-axis representing $K_c$ are from (\ref{eq:Kc_uni}).}
\label{fig:fig4} 
\end{figure}

In this section we consider the Kuramoto model for a uniform distribution $g(\omega)$ for $K$ above the critical 
value, $K>K_c$.  In Fig. (\ref{fig:fig4}) we plot the data points for average value of $r$ as a function of $K$ obtained 
from dynamic simulations.  The results indicate the shift of the curvatures toward lower values 
of $K$ as $\alpha$ increases.  The data points where $K$ goes to zero agree with the theoretical prediction for 
$K_c$ in (\ref{eq:Kc_uni}).  
\graphicspath{{figures/}}

\begin{figure}[h] 
 \begin{center}
 \begin{tabular}{rrrr}
\includegraphics[height=0.21\textwidth,width=0.25\textwidth]{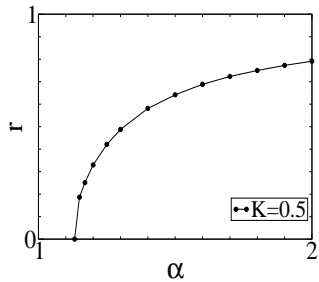} &&
 \end{tabular}
 \end{center} 
\caption{Data points for an average $r$ as a function of $\alpha$ for a uniform $g(\omega)$.  
The system parameters are $\omega_0=1$ and $D=0$.  }
\label{fig:fig4b} 
\end{figure}

A similar plot can be obtained for $r$ plotted as a function of $\alpha$ with fixed $K$, 
see Fig. (\ref{fig:fig4b}).   
\graphicspath{{figures/}}

\section{Conclusion}
\label{sec:5}

In this work we extend the Kuramoto model by incorporating run-and-tumble dynamics.  The extension 
is implemented by an addition of a linear "reaction" term in the Fokker-Planck equation.  On the microscopic 
level, the extension allows individual particles to sample different velocities drawn from the distribution 
$g(\omega)$, where $\alpha$ is the sampling rate.  The original model is recovered when $\alpha=0$, 
in which case individual velocities are fixed.    

How the rate of sampling $\alpha$ affects system dynamics depends on a particular case. For uniform 
and Gaussian distributions, increased $\alpha$ brings about reduced degree of quenched disorder and 
in the limit $\alpha\to\infty$ quenched disorder is completely eliminated and the system behaves as if
$g(\omega)\to\delta(\omega)$.  The reduction of a quenched disorder occasioned by increased $\alpha$
shits down the critical value $K_c$.  

Such a behavior, however, is not universal.  In the case of a 
Lorentz distribution, collective dynamics, or at least the evolution of $r(t)$, is 
independent of $\alpha$, even if the dynamics of individual particles is strongly 
dependent on $\alpha$.  Run-and-tumble dynamics for this distribution does not 
reduce a degree of quenched disorder.  Consequently, $K_c$ is unaffected by 
a sampling rate $\alpha$.

For a discrete bimodal distribution, the situation is also not straightforward as there are 
two types of transitions from an incoherent state.  One of the transitions involves
Hopf bifurcation, in which case the incoherent state transforms to a state with 
$r(t)$ that is periodic in time.  In this case, increased $\alpha$ shifts up the critical 
value of $K$.  If the transformation to a coherent state does not involve Hopf
transformation, then the behavior is similar to that for a Gaussian and uniform 
distribtuions.

\begin{acknowledgments}
D.F. acknowledges financial support from FONDECYT through grant number 1201192.  
D.F.  would like to thank Haim Diamant for introduction to the Kuramoto model.
\end{acknowledgments}

\section{DATA AVAILABILITY}
The data that support the findings of this study are available from the corresponding author upon 
reasonable request.



\end{document}